\documentclass[12pt]{article}
\usepackage{epsfig,graphics}
\newcommand{\be}{\begin{equation}}
\newcommand{\bea}{\begin{eqnarray}}
\newcommand{\eea}{\end{eqnarray}}
\newcommand{\ba}{\begin{array}}
\newcommand{\ea}{\end{array}}
\newcommand{\ee}{\end{equation}}

\expandafter\ifx\csname mathbbm\endcsname\relax

\else

\fi \textheight 22cm \textwidth 15cm \topmargin 1mm \oddsidemargin
5mm \evensidemargin 5mm

\begin{document}
\begin{titlepage}
\hfill \vbox{
    \halign{#\hfil         \cr
           } 
      }  
\vspace*{20mm}
\begin{center}
{\Large {\bf   Nambu-Poisson Bracket and M-Theory Branes Coupled to Antisymmetric Fluxes}
\\}

\vspace*{15mm} \vspace*{1mm} {Mohammad A. Ganjali}
 \\
\vspace*{1cm}

{Department of Fundamental Sciences, Tarbiat Moaallem University\\
Tehran, Iran\footnote {Ganjali@theory.ipm.ac.ir}}

 \vspace*{1cm}
\end{center}
\begin{abstract}
By using the recently proposed prescription \cite{Ho:2008nn} for obtaining the $M5$ brane action from multiple $M2$ branes action in BLG theory, we examine such transition when 11 Dimensional background antisymmetric fluxes couple to the $M2$ brane world volume. Such couplings was suggested in \cite {Li:2008ez} where it was used the fact that various fields in BLG theory are valued in a Lie 3-algebra. We argue that this action and promoting it by Nambu-Poisson bracket gives the expected coupling of fluxes with $M5$ brane at least at weak coupling limit. We also study some other aspects of the action for example, the gauge invariance of the theory.
\end{abstract}

\end{titlepage}

\section{Introduction}
Understanding the structure of $M-$theory as an underlying theory of all known string theories was one of the major efforts during the past decade \cite{Townsend:1996xj}(for a recent review see \cite{Berman:2007bv}). Although several approaches has been found but we don't still have any clear picture of this theory. Fortunately, in the last few years, some ground breaking ideas invented for describing the dynamics of fundamental objects of M theory ie. membranes and five branes.

One of them is a theory proposed by Aharony, Bergman, Jafferis and Maldacena(ABJM) theory\cite{Aharony:2008ug} which is 3D ${\cal N}=6$ superconformal $U(N)\times U(N)$
Chern-Simons gauge theory and describes $N$ coincident $M2$ branes probing $Z_k$ geometry.
The other one is the work of Bagger, Lambert and equivalently Gustavsson(BLG theory)\cite{Bagger:2006sk} where they found a theory for multiple $M2$ branes using a wonderful algebraic structure, Lie 3-algebra\footnote{Recently, a non linear theory for multiple $M2$ branes was proposed in \cite{Iengo:2008cq}.}. The fields in this theory are valued in a non-associative triple algebra as defined by generators $T^a$ as
\bea \label{f1}
[T^a,T^b,T^c]=f^{abc}_{\hspace{.5cm}d}T^d.
\eea
where $f^{abc}_{\hspace{.5cm}d}$ are some constants known as structure constants. By this assumption, one is able to write down a three dimensional ${\cal N}=8$ maximally supersymmetric action which is $U(N)$ gauge theory coupled to Chern-Simons term and is conformal at least at classical level and can explain some expected behavior of multiple membrane dynamics. The proposed action is as

\bea \label{h1}
{\cal L}&=&-\frac{1}{2}<D_{\mu}X_I,D_{\mu}X^I>+\frac{i}{2}<\bar{\Psi},\Gamma^{\mu}D_{\mu}\Psi>\cr
&+&\frac{i}{4}<\bar{\Psi},\Gamma_{IJ}[X^I,X^J,\Psi]>-V(X)+{\cal L}_{CS}
\eea
where
\bea \label{f2}
V(X)=\frac{1}{12}<[X^I,X^J,X^K],[X^I,X^J,X^K]>
\eea
and the Chern-Simons action is
\bea \label{f3}
{\cal L}_{CS}=\frac{1}{2}\epsilon^{\mu\nu\lambda}(f^{abcd}A_{\mu ab}\partial_{\nu}A_{\lambda cd}+\frac{2}{3}f^{cda}_{\hspace{.5cm}g}
f^{efgb}A_{\mu ab}A_{\nu cd}A_{\lambda ef})
\eea
where $X^I,I=1,...,8$ are the transverse coordinates of M2 branes and $\mu,\nu=0,1,2$  specify longitudinal coordinates. Also $a,b,c=1,...,D$ where $D$ is the number of generators of Lie 3-algebra. $D_{\mu}$ is the covariant derivative
\bea \label{f4}
(D_{\mu}\Phi(x))_a=\partial_{\mu}\Phi_a-f^{cdb}_{\hspace{.5cm}a}A_{\mu cd}(x)\Phi_b
\eea
All fields in the theory can be expanded using the generators of the algebra $T^a$ as
\bea \label{f5}
\Phi=\Phi_a T^a
\eea
For closure of the supersymmetry algebra one should impose the following so-called fundamental identity
\bea \label{f6}
f^{cde}_{\hspace{.5cm}g}f^{abg}_{\hspace{.5cm}h}=f^{abc}_{\hspace{.5cm}g}f^{gde}_{\hspace{.5cm}h}
+f^{abd}_{\hspace{.5cm}g}f^{gec}_{\hspace{.5cm}h}+f^{abe}_{\hspace{.5cm}g}f^{cdg}_{\hspace{.5cm}h},
\eea
One also need to define an inner product for generators as $h^{ab}=(\chi^a,\chi^b)$.
The $h^{ab}$ is referred as the metric of the algebra which can have Euclidean or Lorenzian signature.

The fermion field is eleven dimensional Majorana spinor which has $16$ real components and satisfy chirality condition $\Gamma_{012}\Psi=-\Psi$.

After the proposal of BLG, a lot of attempts has been done for extracting and understanding various aspects of this theory, see for example \cite{Ho:2008bn}. One of the important articles in this direction is the work of \cite{Mukhi:2008ux} in which it was shown that if one of the scalars, for example $X^8$, gives an expectation value, one can reduce the membrane action to $D2$ brane action which shows an important notion of reliability of the BLG theory.

Since In M-theory for an $M2$ brane there is an $M5$ brane which are electric-magnetic dual of each other, one natural question is the relation between the $M2$ and $M5$ brane dynamics in the context of the BLG theory. One can use various methods to study $M5$ brane theory by using $M2$ theory \cite{Jeon:2008zj} but the author of \cite{Ho:2008nn} prescribed an interesting approach to achieve this goal using the Nambu-Poisson brackets. In fact, by considering an $3$ dimensional internal space in the world volume of $M2$ brane, they were able to find a six dimensional theory which has some desired properties of an $M5$ brane. For example, they found the action of a self dual two form gauge field living on the world volume of $M5$ brane.

An interesting subject in this theory is the coupling of supergravity form fields as background fields with world volume of M2 and M5 branes and studying their relation via the Ho-Matsuo approach.  Initially, the coupling of antisymmetric fields with $M2$ branes was studied in \cite{Li:2008ez}. It was suggested there a so called Meyers-Chern-Simons(MCS) action, for coupling of antisymmetric form fields $C_3$ and its magnetic dual $C_6$ with $M2$ branes. For rewriting this action the author supposed all fields posses a Lie 3-algebra.

In this paper, we focus on this action for obtaining and studying the $M5$ brane action in the flux sector by using Ho-Matsuo prescription. After a brief introduction of Nambu-Poisson formalism in section 2, we study some aspects such as gauge symmetry of MCS action for $M2$ branes and in the next section, we focus on transition from $M2$ brane theory to $M5$ brane theory. Then, we obtain gauge invariance conditions for $M5$ brane in section 4 and finally in section 5, we study the reduction from $M5$ to $D5$ theory by double dimensional reduction.
\section{$M2$ Brane Coupled to Antisymmetric Fluxes}
The eleven dimensional supergravity contains an antisymmetric three form field $C_3$ and its magnetic dual six form field $C_6$. It is well known that any $p$-brane can covariantly couples to an $p+1$ form field and also to $p-1,...$ form fields due to existence of world volume antisymmetric gauge fields or Kalb-Ramond fields\cite{Polchinski:1995mt}. Meyrs in \cite{Myers:1999ps} showed that $p$-brane can also couple to $p+3,...$ form fields via the fact that the commutators of transverse scalar fields in non-Abelian theories are non zero. Motivating by these and the fact that all fields in the BLG theory can give values in an 3-algebra, the author of \cite {Li:2008ez} suggest the following Myers-Chern-Simons action for the coupling of the $C_3$ and $C_6$ fields to multiple $M2$ branes
\bea \label{f7}
S_{MCS}&=&\lambda_1\int{d^3x \epsilon^{\lambda\mu\nu}C_{IJK}STr(T^aT^bT^c)D_{\lambda}X_a^ID_{\mu}X^J_bD_{\nu}X_c^K}\hspace{.5cm}\cr
&+&\lambda_2\int{d^3x \epsilon^{\lambda\mu\nu}C_{IJKLMN}STr([T^d,T^e,T^f]T^aT^bT^c)X_d^IX_f^JX_f^K
D_{\lambda}X_a^LD_{\mu}X^M_bD_{\nu}X_c^N}
\eea

The "STr" denotes the symmetrized trace and $\lambda_1, \lambda_2$ are some conventional coefficients. One may also define
\bea \label{f8}
g^{abc}=STr(T^aT^bT^c),\hspace{1cm}d^{abcd}=STr(T^aT^bT^cT^d),
\eea
which are symmetric functions due to permutation of indices. In this action, it was ignored the higher order terms involving higher power of Kalb-Ramond field or world volume gauge field. Here, we study some properties of this action as an action of $M2$ brane coupled to antisymmetric fluxes.

Generally, one may consider the background antisymmetric fields as functional of scalar fields $X^I$ and so the MCS terms of action contributes to equation of motion of scalar fields and gauge fields. For example, by computing the variation of the action and noticing that $$\frac{\partial(D_{\lambda}X^I_a)}{\partial(X_m^H)}=-A_{\lambda dc}f^{dcb}_{\hspace{.5cm}m}X^H_b,\;\;\;\;\;\;\;\;\;\;
\partial_{\mu}{(}\frac{\partial(D_{\lambda}X^I_a)}{\partial(\partial_{\mu}X^H_m)})=\partial^2_{\mu}X^H_m$$ one finds the following contribution to the equation of motion of scalar fields
\bea \label{f9}
&+&\lambda_1\epsilon^{\lambda\mu\nu}g^{abc}\frac{\partial(C_{IJK})}{\partial(X_m^H)}
D_{\lambda}X_a^ID_{\mu}X^J_bD_{\nu}X_c^K\cr
&-&3\lambda_1\epsilon^{\lambda\mu\nu}g^{mbc}C_{HJK}A_{\lambda rs}f^{rsh}_{\hspace{.5cm}m}X_h^HD_{\mu}X^J_bD_{\nu}X_c^K\cr
&+&3\lambda_1\epsilon^{\lambda\mu\nu}g^{mbc}C_{HJK}\partial^2_{\lambda}X_m^HD_{\mu}X^J_bD_{\nu}X_c^K\cr
&+&\lambda_2\epsilon^{\lambda\mu\nu}d^{gabc}f^{def}_{\hspace{.5cm}g}
\frac{\partial(C_{IJKLMN})}{\partial(X_m^H)}X_d^IX_e^JX_f^K
D_{\lambda}X_a^LD_{\mu}X^M_bD_{\nu}X_c^N\cr
&+&3\lambda_2\epsilon^{\lambda\mu\nu}d^{gabc}f^{mef}_{\hspace{.5cm}g}C_{HJKLMN}X_e^JX_f^k
D_{\lambda}X_a^LD_{\mu}X^M_bD_{\nu}X_c^N \\
&-&3\lambda_2\epsilon^{\lambda\mu\nu}d^{gmbc}f^{def}_{\hspace{.5cm}g}C_{IJKHMN}A_{\lambda rs}f^{rsh}_{\hspace{.5cm}m}X_d^IX_e^JX_f^K
X_h^HD_{\mu}X^M_bD_{\nu}X_c^N\cr
&+&\lambda_2\epsilon^{\lambda\mu\nu}d^{gmbc}f^{def}_{\hspace{.5cm}g}
C_{IJKHMN}X_d^IX_e^JX_f^K\partial^2_{\lambda}X^H_mD_{\mu}X^M_bD_{\nu}X_c^N
\eea
These terms are, in fact, the higher order terms of equation of motion. Note that there isn't any sum on index $m$.
By varying the action about $A_{\alpha rs}$, the contribution to the gauge field equation of motion is given by
\bea \label{t9}
&&-3\lambda_1\epsilon^{\alpha\mu\nu}g^{abc}C_{IJK}f^{rsh}_{\hspace{.5cm}a}
X_h^ID_{\mu}X^J_bD_{\nu}X_c^K\cr
&&-3\lambda_2\epsilon^{\alpha\mu\nu}d^{gabc}f^{def}_{\hspace{.5cm}g}f^{rsh}_{\hspace{.5cm}a}
C_{IJKLMN}X_d^IX_e^JX_f^K
X_h^LD_{\mu}X^M_bD_{\nu}X_c^N
\eea
One can also find the supersymmetry transformations of MCS terms using the following transformation laws for various fields in the BLG theory
\bea \label{f14}
\delta X_a^I&=&i\bar{\epsilon}\Gamma^I\Psi_a\cr
\delta\Psi_a&=&D_{\mu}X_a^I\Gamma^{\mu}\Gamma_{I}\epsilon-\frac{1}{6}X_b^IX_c^JX_d^Kf^{bcd}_{\hspace{.5cm}a}
\Gamma_{IJK}\epsilon,\cr
\delta\tilde{A}_{\mu\hspace{2mm} a}^{\hspace{2mm}b}&=&\delta (f^{cdb}_{\hspace{.5cm}a}A_{\mu cd})=i\bar{\epsilon}\Gamma_{\mu}\Gamma_{I}C^I\Psi_d f^{cdb}_{\hspace{.5cm}a}
\eea
One finds that the theory is still half BPS as one expects.

Gauge Invariance of the MCS, however, has some new and interesting features because that the gauge invariance of this terms may be preserved in various ways. At the first look, for cancelation of the variation of $C_3$ term in MCS action due to gauge transformation, one may add a boundary term such as
\bea \label{f15}
S_{b}\propto \int_{\partial M_2}{b_2}
\eea
where $b_2$ is a self dual two form field. Then, it can be shown that the action of coupling of the three form field can be canceled. Note that the existence of such self dual fields on the boundary of $M2$ brane can be interpreted as ending of $M2$ branes on $M5$ brane.

Another way is considering the $C_3$ and its dual $C_6$ fields as independent dynamical fields\cite{deAlwis:1997gq} and adding some new terms to the action and restoring the gauge invariance of the full action.

In BLG theory, however, there is another interesting possibility via the fact that the fields in the theory are valued in a Lie 3-algebra. For more details, the gauge transformation of bosonic fields in the BLG theory is given by
\bea \label{f16}
\delta_{\Lambda} X_a^I=\Lambda_{cd}f^{cdb}_{\hspace{.5cm}a}X_b^I,\;\;\;\;\;\;\;\;\;\;\;\delta_{\Lambda}\Psi_a=
\lambda_{cd}f^{cdb}_{\hspace{.5cm}a}\Psi_{b},\cr
\delta_{\Lambda}\tilde{A}_{\mu \hspace{2mm}a}^{\hspace{2mm}b}=\partial_{\mu}\tilde{A}_{\hspace{2mm}a}^b-\tilde{\Lambda}_{\mu\hspace{2mm} c}^{\hspace{2mm}b}\tilde{A}_{\mu \hspace{2mm}a}^{\hspace{2mm}c}+\tilde{A}_{\mu \hspace{2mm}c}^{\hspace{2mm}b}\tilde{\Lambda}_{a}^c\hspace{1.5cm}
\eea
Considering the above transformations we have $\delta_{\Lambda} (D_{\lambda}X_a^I)=\Lambda_{fe}f^{fed}_{a}D_{\lambda}X_d^I$ and
so the variation of MCS action for general form fields is given by
\bea \label{f17}
&&\delta_{\Lambda}({\cal L}_{MCS})=
\lambda_1\epsilon^{\lambda\mu\nu}\{ \delta_{\Lambda} C_{IJK}g^{abc}\cr
&&+\Lambda_{fe}C_{IJK} (g^{dbc}f^{fea}_{\hspace{.5cm}d}+g^{dac}f^{feb}_{\hspace{.5cm}d}
+g^{dab}f^{fec}_{\hspace{.5cm}d})\}D_{\lambda}X_a^ID_{\mu}X^J_bD_{\nu}X_c^K\cr
&&+\lambda_2\epsilon^{\lambda\mu\nu}\{\delta_{\Lambda} C_{IJKLMN}d^{gabc}f^{def}_{\hspace{.5cm}g}
+\Lambda_{ji}C_{IJKLMN}[d^{gabc}f^{jih}_{\hspace{.5cm}g}f^{def}_{\hspace{.5cm}h}\cr
&&+(d^{ghbc}f^{jia}_{\hspace{.5cm}h}+d^{ghac}f^{jib}_{\hspace{.5cm}h}
+d^{ghab}f^{jic}_{\hspace{.5cm}h})f^{def}_{\hspace{.5cm}g}]\}X_d^IX_e^JX_f^K
D_{\lambda}X_a^LD_{\mu}X^M_bD_{\nu}X_c^N,
\eea
where in the third line we have used the fundamental identity. So, for having gauge symmetry for MCS term, one needs to solve following two equations
\bea \label{f18}
&&\delta_{\Lambda}C_{IJK}g^{abc}+C_{IJK}\Lambda_{fe}[g^{dbc}f^{fea}_{\hspace{.5cm}d}+
g^{dac}f^{feb}_{\hspace{.5cm}d}+g^{dab}f^{fec}_{\hspace{.5cm}d}]
=0,\cr
&&\delta_{\Lambda} C_{IJKLMN}g^{abcg}f^{def}_{\hspace{.5cm}g}\\
&&+C_{IJKLMN}\Lambda_{ji}[d^{gabc}f^{jih}_{\hspace{.5cm}g}f^{def}_{\hspace{.5cm}h}+
(d^{ghbc}f^{jia}_{\hspace{.5cm}h}+d^{ghac}f^{jib}_{\hspace{.5cm}h}
+d^{ghab}f^{jic}_{\hspace{.5cm}h})f^{def}_{\hspace{.5cm}g}]=0\nonumber
\eea
For constant background fields, some solutions of these equations are found in \cite{Li:2008ez}. In section $4$ we use the Nambu-Poisson Bracket formalism as a tool for solving these equations.

In the next section, we use this to transfer from $M2$(MCS) to $M5$(MCS) action.
\section{From $M2$ to $M5$}

Finding a $6$ dimensional supersymmetric gauge theory for multiple $M5$ brane is an important and also difficult problems. The difficulty comes from various places. For example, in $6$ dimensional theories there are a self dual antisymmetric two form field in which its dynamics is not very clear. There are several proposal and methods to overcome these difficulties\cite{HST}.

After BLG proposal for multiple $M2$ branes some attempts has been done for studying $M5$ brane theory using the BLG theory\cite{Ho:2008bn}. One of such efforts is the interesting Ho-Matsuo approach\cite{Ho:2008nn}. In their theory, it was supposed that there are internal degrees of freedom in the world volume theory of $M2$ brane. Especially, these internal space is a three dimensional manifold $C(N)$ which one may restrict it by imposing the Nambu-Poisson bracket and then construct a Lie 3-algebra on this manifold(For a review on Nambu-poisson bracket see \cite{Nambu:1973qe,Awata:1999dz}). The whole theory becomes a theory on ${\cal M}\times {\cal N}$ manifold(${\cal M}$ is the world volume of $M2$ brane). In fact, one can see that the physical degrees of freedom are that of a $6$ dimensional ones. In the other hand, there are five transverse directions corresponding to five scalars $X^i$ and $3$ other scalars have the role of three dynamical gauge fields \footnote{Note that in $3$ dimensions the gauge fields are non-dynamical but in $6$ dimensions there are three dynamical gauge fields.}. There is also a 11 dimensional Majorana spinor with a chirality condition which means that it has $16$ real components corresponding to $8$ bosonic degrees of freedom. So, it seems that the theory effectively is a $6$ dimensional theory and $M2$ brane theory promoted to $M5$ brane theory.

In \cite{Ho:2008nn}Ho and Matsuo used this prescription and studied various aspects of $M5$ brane theory. Especially, they found the action of all dynamical fields and their supersymmetry and gauge symmetry transformations. Note that the proposed theory for $M5$ brane is, in fact, a theory for infinite $M2$ branes ending on $M5$ brane\cite{Berman:2006eu}. For finite $M2$ branes one should consider the problem of quantized Nambu-Poisson bracket which is an open problem \cite{Nambu:1973qe,Awata:1999dz}.

In this section, we want to study the coupling of $M5$ brane to supergravity form fields using the above approach.
We consider the internal space ${\cal N}$ is spanned by coordinates $y_{\dot{\mu}}$\footnote{In fact, it is constructed a fiber bundle on {\cal M}\cite{Ho:2008nn}.} and construct a Lie 3-algebra using Nambu-Poisson bracket as
\bea \label{f19}
\{\chi^a,\chi^b,\chi^c\}=\epsilon^{\dot{\mu}\dot{\nu}\dot{\lambda}}
\partial_{\dot{\mu}}\chi^a\partial_{\dot{\nu}}\chi^b\partial_{\dot{\lambda}}\chi^c=f^{abc}_{\hspace{.5cm}d}\chi^d(y)
\eea
where $\chi ^{a}(y)(a=1,...,)$ are the basis of an infinite dimensional space $C(N)$.
The inner product is defined as
\bea \label{f20}
(\chi,\phi)=\int_{N}{d^3y\mu(y)\chi(y)\phi(y)}=\frac{1}{g^2}\int_{N}{d^3y\chi(y)\phi(y)}
\eea
The constant $g$ is the measure factor and is chosen such that the inner product be invariant under the Nambu-Poisson bracket. Any coordinate transformation from $y^{\dot{\mu}}$ to $y'^{\dot{\mu}}=f^{\dot{\mu}}(y)$ should preserves Nambu-Poisson bracket which implies that
\bea \label{f21}
\{f^{\dot{1}},f^{\dot{2}},f^{\dot{3}}\}=1
\eea
and means that these transformations should preserve volume\footnote{So, we do not have full diffeomorphism invariant action on ${\cal N}$ space.}. Note that the definition of metric on manifold ${\cal N}$ is not necessary but the most important constrain is that one should apply only volume diffeomorphism coordinate transformations on this manifold\footnote{It can be shown that the gauge symmetry is volume preserving diffeomorphism.}. Such volume form can be written as $\omega=dy^{\dot{1}}dy^{\dot{2}}dy^{\dot{3}}$.

By using the definition of metric $h^{ab}=(\chi^a,\chi^b)$, one may construct $f^{bcda}=f^{bcd}_{\hspace{.5cm}e} h^{ae}$ which is a totally antisymmetric $4$-tensor.

In this fashion, one can expand all fields of the BLG theory in terms of basis $\chi^a$ of $C(N)$ as
\bea \label{f22}
X^I(x,y)&=&X^I_a(x)\chi ^a(y),\cr
\Psi(x,y)&=&\Psi_a(x)\chi ^a(y),\cr
A_{\mu b}(x,y)&=&X_{\mu ab}(x)\chi ^a(y).
\eea
\subsection{From $M2$ to $M5$ in Flux Sector}
The world volume of an $M5$ brane can naturally couple to an antisymmetric $C_6$ field and also to antisymmetric $C_3$ form field via the existence of a self dual two form field $b_2$ in the world volume of $M5$ brane. The action for these form fields has been studied in several articles (see for example \cite {Townsend:1996xj,deAlwis:1997gq,Douglas:1995bn}). The proposed action can be written as
\bea \label{f23}
S_{M_5}=-\frac{1}{2}\int_{M_5}{{\cal H}_3\wedge *{\cal H}_3}+\int_{M_5}{C_3}\wedge db_2-2\int_{M5}{C_6}
\eea
where the dual field $C_6$ and self dual there form ${\cal H}_3$ is defined such that
\bea \label{f24}
dC_6=*dC-C\wedge dC,\hspace{1.5cm}{\cal H}_3=db_2-C_3|_{M5},
\eea
Note that the definition of $C_6$ is not as usual definition of Hodge star of $C_3$ and also, there is an extra term in the definition of ${\cal H}_3$. In fact, the gauge invariance condition forces such definitions for these form fields. Note also that for obtaining the action (\ref{f23}), one should impose the self duality condition at the level of the equation of motion and then require consistency with Bianchi identity\cite {deAlwis:1997gq}.

Now, for transition from $M2$ brane to $M5$ brane theory, one should find such interacting terms for the coupling of $M5$ brane world volume to these form fields. We will show the existence of such terms in $M5$ brane theory, at weak coupling limit, using the Ho-Matsuo approach.
In this limit, recalling the self duality of $b_2$, we have two terms  $\int_{M_5}{C_3}\wedge *db_2$ and $\int_{M5}{C_6}$  in action (\ref{f23}).

Let us focus on $C_3$ term of MCS action.
Firstly, noting that we deal with an M5 brane theory on ${\cal M}\times {\cal N}$ we have
\bea \label{h24}
\int{C_3 \wedge *db_2}=\int{Pull(C_3)|_{{\cal M}} \wedge *db_2|_{{\cal N}}}+\int{Pull(C_3)|_{{\cal N}} \wedge *db_2|_{{\cal M}}}
\eea
Secondly, we have the following general identity for the wedge product
$$C\wedge *db=db\wedge *C$$
where in its component form can be written as
\bea \label{f32}
\epsilon^{\mu\nu\lambda\dot{\alpha}\dot{\beta}\dot{\gamma}}C_{IJK}{\cal D}_{\mu}X^I{\cal D}_{\nu}X^J{\cal D}_{\lambda}X^K{\cal H}_{\dot{\alpha}\dot{\beta}\dot{\gamma}}=\epsilon^{\dot{\alpha}\dot{\beta}\dot{\gamma}\mu\nu\lambda}
C_{IJK}{\cal D}_{\dot{\alpha}}X^I{\cal D}_{\dot{\beta}}X^J{\cal D}_{\dot{\gamma}}X^K{\cal H}_{\mu\nu\lambda}.
\eea
Then the covariant derivatives of a covariant field $\Phi$ can be defined as\cite {Ho:2008nn}
\bea \label{t33}
{\cal D}_{\underline{\mu}}\Phi=\partial_{\underline{\mu}}\Phi-g\{b_{\underline{\mu}\underline{\nu}},y^{\underline{\nu}},\Phi \}+\frac{g^2}{2}\epsilon_{\underline{\mu}\underline{\nu}\underline{\rho}}\{b^{\underline{\mu}},b^{\underline{\nu}},\Phi \}
\eea
where $y^{\underline{\mu}}$ is the collective coordinate for $y^{\mu}$ and $y^{\dot{\mu}}$. Note that because of the definition of Nambu-Poisson bracket the third term trivially is equal to zero whenever $\underline{\mu}=\dot{\mu}$. Equation (\ref{f32}) means that
\bea \label{h25}
\int{C_3 \wedge *db_2}=2\int{d^6\underline{y}\epsilon^{\mu\nu\lambda\dot{\alpha}\dot{\beta}\dot{\gamma}}C_{IJK}{\cal D}_{\mu}X^I{\cal D}_{\nu}X^J{\cal D}_{\lambda}X^K{\cal H}_{\dot{\alpha}\dot{\beta}\dot{\gamma}}}
\eea
So, we need the following equality
\bea \label{h26}
\lambda_1\epsilon^{\mu\nu\lambda}C_{IJK}{\cal D}_{\mu}X^I{\cal D}_{\nu}X^J{\cal D}_{\lambda}X^K=2\epsilon^{\mu\nu\lambda\dot{\alpha}\dot{\beta}\dot{\gamma}}C_{IJK}{\cal D}_{\mu}X^I{\cal D}_{\nu}X^J{\cal D}_{\lambda}X^K{\cal H}_{\dot{\alpha}\dot{\beta}\dot{\gamma}}
\eea
\newpage
This equality can be satisfied if one chooses a Lorenz-like gauge
\bea \label{h27}
\partial_{\dot{\rho}}b^{\dot{\rho}}=g'
\eea
where $g'$ is a nonzero constant. By an appropriate choice of $\lambda_1$ and $g'$ one can recover (\ref{h26}). In this gauge one may obtain the equation for gauge parameter as
$$\epsilon^{\dot{\mu}\dot{\nu}\dot{\rho}}\partial_{\dot{\rho}}b^{\dot{\alpha}}\partial_{\dot{\alpha}}
\partial_{\dot{\mu}}\Lambda_{\dot{\nu}}=0$$

In the next step, we analyze the $C_6$ term in MCS action. Here again we deal with the pull back of six form on $M5$ brane world volume which is ${\cal M}\times {\cal N}$ and so one may decompose $C_6$ as $C^1_3\wedge C^2_3$ and then obtain the pull back of these fields on ${\cal M}\times {\cal N}$.
Using the definition of Nambu-Poisson bracket one obtains (\ref{f7})
\bea \label{f25}
\lambda_2\int{d^3xd^3y\epsilon^{\lambda\mu\nu\dot{\lambda}\dot{\mu}\dot{\nu}}C_{IJKMNP}
\partial_{\dot{\lambda}}X^I\partial_{\dot{\mu}}X^J\partial_{\dot{\nu}}X^K
D_{\lambda}X_a^LD_{\mu}X^M_bD_{\nu}X_c^N}
\eea
This term is expected to be the interacting term of $C_6$ field to the world volume of $M5$ brane but, it is the weak coupling limit of such coupling. In fact, the Nambu-Poisson bracket is not defined in terms of covariant derivatives of the fields and we have non covariant derivatives in (\ref{f25}). But in the weak coupling $g\rightarrow 0$ we have ${\cal D}_{\dot{\mu}}\simeq \partial_{\dot{\mu}}$. For obtaining the higher order terms of the coupling of $C_6$ with world volume of the M5 brane one should consider all higher order terms in MCS action.

Note that although the terms $\partial_{\dot{\mu}}$ are appear non-covariantly but, the full action (\ref{f25}) is covariant. In fact, the triple product of covariant fields is also covariant because of the fundamental identity\cite{Ho:2008nn}. In the other word, if $\Phi_a,a=1,2,3$ are some covariant Lie 3-algebra valued fields then $\{\Phi_1,\Phi_2,\Phi_3\}$ is also covariant because that the gauge transformation for any covariant field in BLG theory is given by
\bea
\delta_{\Lambda}\Phi=\Lambda_{ab}\{T^a,T^b,\Phi\}.
\eea
By considering
\bea \label{f26}
\delta\{\Phi_1,\Phi_2,\Phi_3\}=\{\delta\Phi_1,\Phi_2,\Phi_3\}+
\{\Phi_1,\delta\Phi_2,\Phi_3\}+\{\Phi_1,\Phi_2,\delta\Phi_3\}
\eea
and recalling the fundamental identity (\ref{f6}) one finds that $\{\Phi_1,\Phi_2,\Phi_3\}$  also transforms covariantly.
So, the above action is the covariant form of the pull back of antisymmetric $C_6$ on $M5$ brane world volume at the weak coupling.

In summary, we obtain the desired terms of the coupling of the background antisymmetric fluxes with the world volume of $M5$ branes at least at weak coupling limit $g\rightarrow 0$.
Other terms correspond to higher order terms in the $M5$ brane action which is believed that can be written in a DBI type action\cite{Iengo:2008cq}.

Note that, the Nambu-Poisson bracket and fundamental identity of Lie 3-algebra have crucial rule in derivation of $M5$ brane action from $M2$ brane action.

\section{$M5$ Brane Coupled to Antisymmetric Fluxes}
In this section, we point out some aspects of the coupling of fluxes with $M5$ brane world volume. As usual, one can fine the equation of motion of various fields when these fluxes exist. For example, one may find
\bea
d{\cal H}_3=-dC_3|_{M5}
\eea
which in fact is the Bianchi identity. One can also find the equation of motions in the component form in general cases where the form fields are functional of transverse fields $X^i$'s.

Besides, the field contents of an $M5$ barne, the self dual three form field ${\cal H}_3$, chiral fermion field $\Psi'$\footnote{This fermion field is defined in terms of BLG theory fermion field $\Psi$ as $\Psi=\frac{1}{\sqrt{2}}(1-\Gamma_{\dot{1}\dot{2}\dot{3}})\Psi'.$}, and five scalar fields $X^i$, form a tensor multiplet of ${\cal N}=(0,2)$ supersymmetry\cite{Ho:2008nn}. In the presence of fluxes the theory is still $\frac{1}{2}BPS$ but the unbroken sector of the theory changes.

For the gauge invariance of the theory, one may be interested on solving the equations (\ref{f18}) without using the usual gauge invariance conditions in $M$ theory. By using the definition of Nambu-Poisson bracket, one can show that, in terms of the $C(N)$ basis, these two equations reduces to following equations
\bea \label{f33}
\Lambda_{ij}\{\chi^i,\chi^j,\chi^a\chi^b\chi^c\}=0,\hspace{.4cm}\\
\Lambda_{ij}\{\chi^i,\chi^j,\{\chi^d,\chi^e,\chi^f\}\}=0
\eea
Note that here the $STr$ function is just the integral of basis over ${\cal N}$ space.
In fact, the above equations are a set of differential equations for $C(N)$ basis. The solutions for these equations, if any exist, guarantee  the gauge invariance of the action. As we mentioned before, some non trivial solutions are found in \cite{Li:2008ez} for finite dimensional algebra, However, in some cases one may not find solutions for these equations or find trivial solutions. Here we present an example.

Consider $T^3$ manifold with radius $R$. The basis of functions can be parameterized by $\vec{n}\in Z^3$ as \cite{Ho:2008nn}
\bea \label{f34}
\chi^{\vec{n}}(\vec{y})=\exp{2\pi i \vec{n}.\vec{y}/R}
\eea
So, the metric and structure constants can be written by
\bea \label{f35}
h^{\vec{n_1}\vec{n_2}}&=&\delta(\vec{n_1}+\vec{n_2}),\cr
f^{\vec{n_1}\vec{n_2}\vec{n_3}}_{\hspace{1cm}\vec{n_4}}&=&(2\pi i/R)^3\vec{n_1}.(\vec{n_2}\times\vec{n_3})
\delta(\vec{n_1}+\vec{n_2}+\vec{n_3}-\vec{n_4})
\eea
Note that the measure is chosen $\mu=(2\pi R)^{-3}$. In $R\rightarrow\infty$ limit, one obtains the Lie 3-algebra for $R^3$ where the label of basis becomes continuous and the metric becomes delta function. Using these basis, the gauge invariance conditions read as
\bea \label{f36}
\vec{a}+\vec{b}+\vec{c}=0,\hspace{2.1cm}\cr
\vec{d}+\vec{e}+\vec{f}=0\hspace{.8cm}or\hspace{.8cm}\vec{d}.(\vec{e}\times \vec{f})=0
\eea
which means that we have trivial solutions
\bea
f^{abc}_{\hspace{.5cm}d}=0
\eea
So, one can not restore the gauge invariance of the action by solving the equations (\ref{f36}) and
one should use the full action (\ref{f23}) which is invariant under the usual supergravity gauge transformations.
\section{From $M5$ to $D4$}
As it was shown in \cite{Ho:2008nn}, one can reduce the $M5$ brane action by double dimensional reduction of the six dimensional theory and obtain the action of non-commutative $U(1)$ gauge theory on a $D4$ brane in a $B$-field background. So, one may apply such double dimensional reduction in the sector of coupling of fluxes with the world volume of $M5$ brane. Since this reduction gives an effective theory for $D4$ brane in $IIA$ string theory, one expects that the reduced action shows the coupling of background $C_5$ form field with $D4$ brane world volume as $\int_{D4}{C_5}$ and background $C_3$ form field with $D4$ brane world volume via the two form gauge field on $D4$ brane as $\int_{D4}{C_3\wedge H_2}$. These can be shown easily as follows.

In double dimensional reduction one wraps one leg of the M5 brane on a compactified dimension(for example $X^{\dot{3}}$) and supposes all fields except $X^{\dot{3}}$ are independent of $y^{\dot{3}}$ i.e
\bea
X^{\dot{3}}=\frac{1}{g}y^{\dot{3}}
\eea
and imposes a periodicity condition $X^{\dot{3}}\sim X^{\dot{3}}+L_{11}$. By the gauge invariance one can set $b^{\dot{3}}=0$. By the above assumptions, one easily obtains $\partial_{\dot{3}}y^{\underline{\nu}}=\delta_{\dot{3}}^{\underline{\nu}}$ therefore the three form ${\cal H}_{\dot{\alpha}\dot{\beta}\dot{\gamma}}$ effectively reduces to a two form field $H_{\dot{\alpha}\dot{\beta}}$ and so
\bea
\int_{M5}{C_3\wedge *{\cal H}_3}\sim \int_{D4}{C_3\wedge H_2}
\eea
which is the coupling of three form with $D4$ brane via the world volume two form field.

Similarly, for the $C_6$ term we have
\bea
\lambda_2\int{d^3xd^3y\epsilon^{\lambda\mu\nu\dot{\lambda}\dot{\mu}\dot{\nu}}g^{abc}C_{IJKLMN}
\partial_{\dot{\lambda}}X^I\partial_{\dot{\mu}}X^J\partial_{\dot{\nu}}X^K
D_{\lambda}X_a^LD_{\mu}X^M_bD_{\nu}X_c^N}\cr \sim \lambda_2\int{d^3xd^3y\epsilon^{\lambda\mu\nu\dot{\mu}\dot{\nu}\dot{3}}g^{abc}C_{\dot{3}JKLMN}
\partial_{\dot{\mu}}X^J\partial_{\dot{\nu}}X^K
D_{\lambda}X_a^LD_{\mu}X^M_bD_{\nu}X_c^N}\hspace{.6cm}
\eea
which is the coupling of $C_5$ form field with world volume of $D4$ brane.

Note that $D4$ brane can also couples to background one form field of IIA theory via the higher power of two form field $H_2$ which we did'nt consider such higher order couplings here.
\section{Conclusion}
By using the concepts and structures of Nambu-Poisson manifolds, Ho and Matsuo\cite{Ho:2008nn} were able to promote the three dimensional field theory on multiple $M2$ branes to a six dimensional ones which is realized as world volume theory on an $M5$ brane. The theory has field contents of ${\cal N}=(0,2)$ supersymmetric theory ie. five scalars $X^i$, the self dual tensor field ${\cal H}_3$ and chiral fermion field $\Psi'$ which all of them are valued in a Lie 3-algebra. The corresponding supersymmetry and gauge symmetry transformations of fields is derived too.

It is natural to demand that such prescription works when the 11 dimensional supergravity antisymmetric fields(as background fields) couple to membrane. In this paper, we argue that at least in the weak coupling limit of gauge theory it happens and one can see such transition from $M2$ brane to M5 baren theory. Our starting point is the action presented in \cite{Li:2008ez} where it has been suggested a so called Myers-Chern-Simon(MCS) action for the coupling of the $M2$ brane to background antisymmetric fluxes $C_3$ and $C_6$. The main ingredient in this action is that it is supposed that all fields are valued in a Lie 3-algebra. After applying the Ho-Matsuo approach to this action, we find the term which, in a Lorenz-like gauge, is the coupling of $M5$ brane to $C_3$ and the coupling of M5 to $C_6$ at weak coupling limit. It is important to study the behavior of MCS term in strong coupling limit.

We also study some other aspects of M5 brane in the flux sector especially, we obtain a set of differential equations which their solutions are sufficient conditions for having gauge invariance in the theory. Then, for solving these equations, we present an example, $T^3$ manifold, and apply this conditions on this space. We find that one can not find a non trivial solutions for these equations and so, for having gauge invariance one should add some extra terms(boundary terms or higher order terms) to restoring gauge invariance.

At the end, it also seems that the Lie 3-algebra has an important roll in transition from $M2$ brane theory to $M5$ brane theory. It may also has a roll in a duality in $M$ theory known as $M2-M2$ brane duality, which in some simple form is a duality between membrane and five brane wrapped on a three dimensional manifold, which was originally proposed in \cite{Townsend:1995de}(See also \cite{Aharony:1996wp}).

\section{Acknowledgment}
I would like to thank Dr A. Ali-Akbari for drawing my attention to the BLG theory.

\end{document}